\documentstyle{mn}
\input epsf
\begin{document}
\newcommand{\mincir}{\ \raise -2.truept\hbox{\rlap{\hbox{$\sim$}}\raise5.truept
	\hbox{$<$}\ }}			
\newcommand{\magcir}{\ \raise -2.truept\hbox{\rlap{\hbox{$\sim$}}\raise5.truept
 	\hbox{$>$}\ }}		
\newcommand{\refer}{\par\noindent\hangindent 20pt}
\def\OVI{\hbox{O~$\scriptstyle\rm VI\ $}}
\def\HI{\hbox{H~$\scriptstyle\rm I\ $}}
\def\HII{\hbox{H~$\scriptstyle\rm II\ $}}
\def\CIV{\hbox{C~$\scriptstyle\rm IV\ $}}

\title[THE STELLAR UV BACKGROUND]{THE STELLAR UV BACKGROUND AT z$<$1.5 AND THE 
BARYON DENSITY OF PHOTOIONIZED GAS}

\author[E. Giallongo et al.]
{
E. Giallongo$^1$, A. Fontana$^1$, P. Madau$^2$\\
$^1$ Osservatorio Astronomico di Roma, I-00040, Monteporzio, Italy\\
$^2$ Space Telescope Science Institute, 3700 San Martin Drive, Baltimore
MD 21218
}
\maketitle

\begin{abstract}

We use new studies of the cosmic evolution of star-forming galaxies to
estimate the production rate of ionizing photons from hot, massive
stars at low and intermediate redshifts. The luminosity function of
blue galaxies in the Canada-France Redshift Survey shows appreciable
evolution in the redshift interval $z=0-1.3$, and generates a
background intensity at 1 ryd of $J_L\approx 1.3\times 10^{-21}\langle
f_{\rm esc}\rangle$ ergs cm$^{-2}$ s$^{-1}$ Hz$^{-1}$ sr$^{-1}$ at
$z\approx 0.5$, where $\langle f_{\rm esc}\rangle$ is the unknown
fraction of stellar Lyman-continuum photons which can escape
into the intergalactic space, and we have assumed that the absorption is 
picket fence-type. We argue that recent upper limits on the
H$\alpha$ surface brightness of nearby intergalactic clouds constrain this
fraction to be $\mincir 20$\%. The background ionizing flux from galaxies can
exceed the QSO contribution at $z\approx 0.5$ if $\langle f_{\rm
esc}\rangle\magcir 6\%$. 
We show that, in the general framework of a diffuse background
dominated by QSOs and/or star-forming galaxies, the cosmological
baryon density associated with photoionized, optically thin gas
decreases rapidly with cosmic time. The results of a recent {\it
Hubble Space Telescope} survey of \OVI absorption lines in QSO spectra
suggest that most of this evolution may be due to the bulk heating and
collisional ionization of the intergalactic medium by supernova events
in young galaxy halos.
\end{abstract}
\nokeywords
\section{Introduction}
The integrated ultraviolet flux arising from quasars and hot, massive
stars in star-forming galaxies is likely responsible for maintaining
the high degree of ionization of the intergalactic medium (IGM).  The
large, low-density intergalactic clouds which produce the
Lyman-$\alpha$ forest lines in the absorption spectra of background
QSOs represent one of the observational signature of this high level
of ionization (Bechtold 1994; Giallongo et al. 1996).  Locally, the
ultraviolet ionizing background (UVB) may be responsible for the
ionization of the hydrogen clouds located in the galactic halo
(Ferrara \& Field 1994), and for producing the abrupt truncation in
the H~I distribution at the edge of nearby spiral galaxies (Bochkarev
\& Sunyaev 1977).

The contribution of QSOs to the UVB is the easiest to assess with some
degree of reliability. Madau (1992), and, more recently, Haardt \&
Madau (1996) have computed the intensity of the UVB as a function of
redshift based on the most recent estimates of the quasar luminosity
function from $z=0$ to $z\approx 5$.  They included the reprocessing
of ultraviolet radiation by intergalactic material, and showed that
QSO absorption-line systems are sources, not just sinks of ionizing
photons. The resulting metagalactic flux at the Lyman edge is found to
increase from $\approx 10^{-23}$ ergs cm$^{-2}$ s$^{-1}$ Hz$^{-1}$
sr$^{-1}$ at the present-epoch to $\approx 5\times 10^{-22}$ ergs
cm$^{-2}$ s$^{-1}$ Hz$^{-1}$ sr$^{-1}$ at $z=2.5$. The background
intensity stays nearly constant in the redshift interval $z=1.5-3.5$,
to drop rapidly beyond $z=4$ because of the steep decline of the
quasar population.

Hot, massive stars in star-forming galaxies have also been suggested
as important contributors to the UVB (Bechtold et al. 1987;
Songaila, Cowie, \& Lilly 1990; Miralda-Escud\'e \& Ostriker 1990) at
early epochs.  As the present rate of production of metals in normal
galaxies is too low to account for the observed element abundances,
there must have been an epoch when the heavy element production rate
per unit mass was several times larger than it is today (Madau et
al. 1996). Data on the galaxy population at $z\magcir 3$ are sparse
but are accumulating rapidly (Steidel et al. 1996; Madau et
al. 1996). The recent detection by Tytler et al. (1995) and by Cowie
et al. (1995) of \CIV in the Lyman-$\alpha$ forest clouds has provided
the first evidence of widespread chemical enrichment in the IGM at
$z\sim 3$. Madau \& Shull (1996) have computed the ionizing stellar
radiation flux which accompanies the production of metals at high-$z$,
and found that this may be significant, comparable to the QSO
contribution if a fraction $\magcir 25$\% of the UV radiation emitted
from stars can escape into the intergalactic space.  At low and
intermediate redshifts, the Canada-France Redshift Survey (Crampton et
al. 1995) has provided new information on the properties and evolution
of field galaxies at $z<1.3$. Finally, the first survey for \OVI 1032,
1038\AA\ absorption lines in QSO spectra (Burles \& Tytler 1996) has
shown the likely presence of a substantial cosmological mass density
of hot, collisionally ionized gas at $\langle z\rangle =0.9$.

In this paper we estimate the contribution of blue, star-forming
galaxies to the UVB, and use some new upper limits to the intensity of
the UV diffuse radiation field at the present-epoch to constrain the
average escape fraction of Lyman-continuum (Lyc) photons from such
systems. We also show that the cosmological gas density of $T\approx
20,000$ K material photoionized by the UVB decreases strongly with
cosmic time, and suggest that the recent detection of \OVI
intergalactic absorption lines may provide some clues on the fate of
the ``missing'' baryons. Throughout this paper, we shall adopt a flat
cosmology with $H_0=50$h$_{50}$ km s$^{-1}$ Mpc$^{-1}$ and $q_0=0.5$.

\section{STAR-FORMING GALAXIES AND THE UVB AT z$<$1.5}
\subsection{Blue Galaxy Emissivity}

\bigskip
The results of the Canada-France Redshift Survey have made it possible
to disantangle the evolution of the luminosity function (LF) from
redshift $z=0.02$ to redshift $z=1.3$ of the blue and red galaxy
population separately (Lilly et al. 1995).  At variance with that of
red objects, the LF of blue galaxies (bluer than present day Sbc)
shows significant cosmological evolution. This evolution can be
represented as a brightening of the average luminosity of the blue
population as the redshift increases (although density evolution can
be present as well). Between $z\approx 0.3$ and $z\approx0.6$ the LF
brightens by about 1 magnitude. Beyond, and up to $z\approx 1.3$ the
luminosity evolution of the bright end of the LF levels off. For
$z<0.2$, the new survey shows a significant excess relative to the
local LF of Loveday et al. (1992).

In the AB photometric system, the average rest-frame ultraviolet color
of the blue population, in the redshift interval considered, is
$\langle U-V \rangle\approx 0.8$. To estimate the mean number of Lyc
photons produced by O-B stars, we have modeled the intrinsic UV
spectral energy distribution of star-forming galaxies using the
Bruzual \& Charlot (1993) stellar population synthesis code. We assume
a Salpeter initial mass function with lower and upper cutoffs of 0.1
and 125$M_{\odot}$, and a constant star formation rate.  The fiducial 
galaxy age is fixed at $\approx 2$ Gyr; this yields the required $U-V$ color
after convolving with the standard Johnson broadband filters. Differences
greater than 1 Gyr in the age produce changes greater than 0.1 mag in the
average colors.

The comoving emissivity at $h\nu_L=1\,$ryd of our galaxy sample can then be
written as \begin{equation} \epsilon (\nu_L,z)=\langle f_{\rm esc}\rangle
{E_{\nu_L}\over E_{\nu _B}}\int_{L_{min}}^{\infty} \phi(L,z) L dL,
\end{equation} where $\langle f_{\rm esc}\rangle$ is the fraction of Lyc
photons emitted from stars which can escape into the intergalactic space, 
assumed to be independent of frequency (picket fence-type absorption),
in analogy with the leakage of ionizing photons from the galactic disk.
$E_\nu$ is the spectral energy distribution. The luminosity function in the
B-band, $\phi$, at various $z$ is taken from Lilly et al. (1995), and we have
adopted a constant value of $L_{min}$, corresponding to $M_B=-17.8$. The
ensuing galaxy emissivity at the Lyman edge is plotted in Figure 1 as a
function of redshift: a strong cosmological evolution is apparent from $z=0$ to
$z\approx 1$. 

Our procedure yields $\epsilon (912$\AA$,0)=2\times 10^{25} \langle
f_{\rm esc}\rangle$ ergs s$^{-1}$ Hz$^{-1}$ Mpc$^{-3}$.  To check for
systematic errors, we have re-computed the present-day ionizing emissivity
starting from the H$\alpha$ luminosity density of the local universe, $1.3\pm
0.7\times 10^{39}$ ergs s$^{-1}$ Mpc$^{-3}$ (Gallego et al. 1995). Assuming
case-B recombination theory, an escape fraction of 50\%, and our fiducial
population synthesis galaxy spectrum, this value implies $\epsilon
(912$\AA$,0)=1.1\pm 0.5\times 10^{25}$ ergs s$^{-1}$ Hz$^{-1}$ Mpc$^{-3}$, in
good agremeent, within the errors, with the extrapolated value plotted in
Figure 1. 

\begin{figure}
{~
\epsfxsize=250 pt
\epsffile{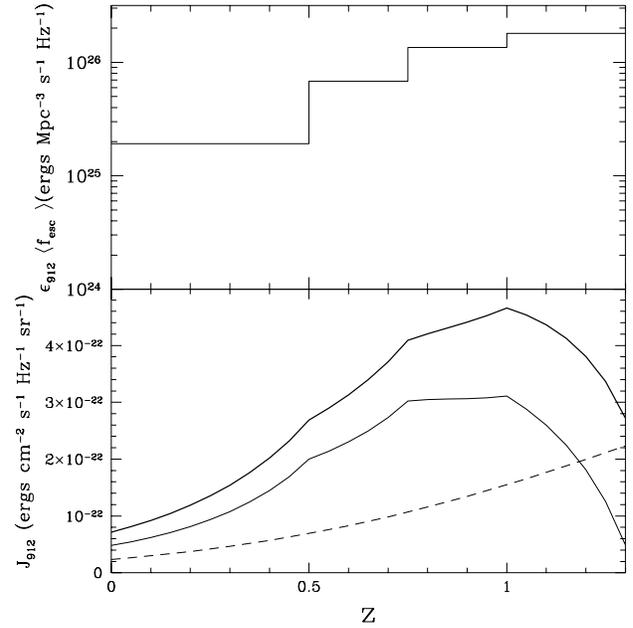}
}
\caption{
{\it a)} Galaxy emissivity as a function of redshift derived
from the binned luminosity function of Lilly et al. (1995).
 {\it b)} The UV background as a
function of redshift produced by galaxies ({\it solid thin curve}), quasars 
({\it dashed curve}), and galaxies $+$ quasars ({\it solid thick curve}). 
The galaxy contribution has been computed assuming an escape fraction of
ionizing photons into the IGM of $\langle f_{esc} \rangle=15$\%. 
}
\label{fig1}
\end{figure}

\subsection{Ionizing Background Intensity}

The mean specific intensity $J_{\nu}$ of the UV background at the observed
frequency $\nu_{o}$, as seen by an observer at redshift $z_{o}$, is given by
\begin{equation}
J(\nu_o,z_o)={1\over 4\pi}\int_{z_o}^\infty dz {d\ell \over dz}{(1+z_o)^3\over
(1+z)^3}\epsilon(\nu,z)\exp[-\tau_{eff}(\nu_o,z_o,z)]
\end{equation}
where $\tau_{eff}$ is the effective photoelectric optical depth of a cloudy
intergalactic medium following Madau (1992), $\epsilon (\nu,z)$ is the proper
volume emissivity at frequency $\nu=\nu _o(1+z)/(1+z_o)$, and $d\ell /dz$ is
the line element in a Friedmann cosmology. 

The main uncertainty in the estimate of the UVB generated by
star-forming galaxies is the escape fraction of Lyc photons $\langle
f_{\rm esc}\rangle $ into the IGM. 
While theoretical models of the structure of \HII regions in a
vertically stratified disk suggest an escape fraction of $\sim 15$\%
through the \HI layers into the halo (Dove \& Shull 1994), the recent
data obtained by the {\it Hopkins Ultraviolet Telescope} show that
less than 3\% of the stellar UV radiation can escape from the four,
low-redshift starburst galaxies observed by Leitherer et
al. (1995). Given the peculiar nature of the selected targets,
however, it is conceivable that the escape fraction might be
significantly reduced in these objects by the presence of dust
absorption. From a reevaluation of the same data Hurwitz
et al. (1997) suggest less restrictive upper limits with an average
value closer to $19$\%.

An indirect argument, put forward by Patel \& Wilson (1995a, b),
suggests a Lyc escape fraction from the galactic disk into the halo
greater than 50\%.  From their observations of H$\alpha$ emission in
nearby galaxies, these authors were able to estimate the total star
formation rate and compare it with the value derived from the
distribution of OB stars in the galactic disk.  In the presence of
ionization balance, the two values should be consistent. On the
contrary, heavy losses of ionizing photons appear to be taking place
in the field. In this respect, it is important to note that, although
the fraction of ionizing photons leaking from the galactic disk could
be relatively high, Lyc absorption by cold gas in galaxy halos may be
quite common, as suggested by the identification of the galaxies
associated with the Lyman-limit systems observed in quasar spectra
(e.g., Steidel 1995).

The study of nearby isolated intergalactic clouds where no sign of
stellar activity is detected can provide a stringent upper limit to
the local UVB and hence to the mean escape fraction. In particular, by
looking at the H$\alpha$ radiation that should escape from these
optically thick clouds due to the reprocessing of the incident Lyc
photons, Vogel et al. (1995) (see also Donahue, Aldering \& Stocke
1995) have recently set a 3-$\sigma$ upper limit of $J_L<8\times
10^{-23}$ ergs cm$^{-2}$ s$^{-1}$ Hz$^{-1}$ sr$^{-1}$ at $z=0$.  Given
the constraints on the luminosity function and the average spectral
shape of the blue galaxies, the local UVB limit implies that no more
than $20$\% of Lyc photons can escape into the IGM from star-forming
regions. The evolution of a galaxy-dominated UVB is shown as a
function of redshift in Figure 1. Adding together the galaxy
contribution with the one estimated from quasars and local AGNs
(Haardt \& Madau 1996), we find that a value of $\langle f_{\rm esc}
\rangle =15$\%, predicted by theoretical models (Dove \& Shull 1994),
yields a total UVB which is still consistent with the local upper
limits. It is interesting to notice that, since the opacity of the IGM
at $z<1.5$ is low, galaxies at intermediate redshifts can provide a
non negligible contribution to the local metagalactic flux. The
current limits on the UVB from H$\alpha$-brightness studies in fact
provide the most stringent constraints to the total ionizing
emissivity at $z\approx 0.5-1$, where the bulk of the blue galaxy
evolution is observed.

As shown in Figure 1, an escape fraction of $\langle f_{\rm esc}
\rangle =15$\% implies a total UVB of $J_L\sim 10^{-22}$ ergs
cm$^{-2}$ s$^{-1}$ Hz$^{-1}$ sr$^{-1}$ at $z\simeq 0.15$. 
If the contribution of the
blue galaxies to the metagalactic flux extends to $z\mincir 1.5$,
i.e. if the escape fraction of Lyc photons into the IGM is significant
even at early epochs, it is possible to envisage a scenario where the
background intensity remains nearly constant, $J_L=2-4\times 10^{-22}$
ergs cm$^{-2}$ s$^{-1}$ Hz$^{-1}$ sr$^{-1}$, from $z=0.4$ to $z=4$,
i.e. for a fraction of 50\% of the age of the universe.
Alternatively, QSOs will dominate the metagalactic flux if $\langle
f_{\rm esc} \rangle<6$\%. The intensity,
spectrum, and evolutionary history of a QSO-dominated UVB have been
recently computed by Haardt \& Madau (1996) on the basis of our
current knowledge of the QSO luminosity function. The integrated
diffuse flux at the Lyman edge is found to increase from $2\times
10^{-23}$ at $z=0$ to $\approx 5\times 10^{-22}$ ergs cm$^{-2}$
s$^{-1}$ Hz$^{-1}$ sr$^{-1}$ at $z\approx 2$. Over the redshift range
$z=2-3.5$, $J_L$ remains roughly constant, to decrease to $2\times
10^{-23}$ ergs cm$^{-2}$ s$^{-1}$ Hz$^{-1}$ sr$^{-1}$ at $z\approx 5$.

Since the luminosity function of star forming galaxies at $z>1$ is
poorly known it is difficult to estimate the galaxy contribution to
the UVB in the redshift interval $2<z<5$. However an extrapolation
of the $z\simeq 1$ luminosity function at higher $z$ would imply
a total UVB at $z\simeq 3.5$ of the order of 
$J_L=6\pm 1\times 10^{-22}$ ergs cm$^{-2}$ s$^{-1}$ Hz$^{-1}$
sr$^{-1}$.

An independent estimate of $J_L$ at high redshifts comes from the
statistical study of the proximity effect in the spectra of QSOs
(Bajtlik, Duncan, \& Ostriker 1988; Bechtold 1994). The value recently
derived from a large high resolution sample of Lyman-$\alpha$ forest
lines is $J_L=5\pm 1\times 10^{-22}$ ergs cm$^{-2}$ s$^{-1}$ Hz$^{-1}$
sr$^{-1}$, constant in the redshift range $1.7<z<4.1$ (Giallongo et
al. 1996). As a direct consequence, while in a QSO-dominated
background model the photoionization rate will remain approximately
constant from $z\sim 3.5$ to $z\sim 1.5$, to drop by a factor of $\sim
30$ by the present epoch, in a universe in which star-forming galaxies
contribute significantly to the metagalactic flux, $J_L$ may only
decrease by a factor $\sim 6$ from $z=2$ to the present epoch.

\section{THE BARYON DENSITY OF PHOTOIONIZED GAS}

A simple estimate of the cosmological density of the baryons hidden in
the Lyman-$\alpha$ forest clouds can be obtained from the following
quantities: the intensity $J_L$ of the UVB, the number density of
lines along the line of sight ${dN/dz}$, the typical cloud radius $R$,
and the observed \HI column. The latter spans a large range of values,
from $10^{12}$ cm$^{-2}$ to $10^{15}$ cm$^{-2}$ and more, showing a
bending below $N_{\rm HI}\approx 10^{14}$ cm$^{-2}$ where the slope of
the power-law distribution flattens from 1.8 to 1.4 at $\langle
z\rangle = 3$ (Giallongo et al. 1996). The density of lines with
$N_{\rm HI}\magcir 10^{14}$ cm$^{-2}$, as derived from {\it Hubble
Space Telescope} (Bahcall et al. 1996) and optical data (Giallongo et
al. 1996), stays approximately constant -- $ dN/dz\sim 25$ -- from
$z\approx 0.15$ to $z\approx 1.8$, to increase rapidly for $z>2$, $
dN/dz\sim 220$ at $z\approx 3.8$.

The baryon density of photoionized gas can be expressed in units of
the critical value $n_{\rm crit}$ as $\Omega_{\rm IGM}= f_V n_{\rm
H}/n_{\rm crit}$ where $f_V$ is the volume filling factor and $n_{\rm
H}$ is the typical hydrogen gas density of individual clouds. The
volume filling factor is proportional to the number of systems along
the line of sight and depends on the cloud geometry, while the gas
density is a measure of the photoionization state of the absorbers
(Madau \& Shull 1996), $n_{\rm H}\propto R^{-1/2}N_{\rm HI}^{1/2}
T^{0.363} J_L^{1/2}(a~\cos\theta) ^{1/2}$.  The temperature dependence
is very weak for highly photoionized clouds: a value of $T=2\times
10^4$ K has been adopted here. The aspect ratio $a\equiv R/l$
generalizes the absorber geometry from spheres to disks of transverse
radii R to half-thickness $l$. For $\langle a~\cos \theta \rangle =1$
the usual limit of spherical clouds is obtained.

Hence, the cosmological mass density produced by the Ly$\alpha$ clouds
is proportional to
\begin{equation}
\begin{array}{ll}
\Omega_{\rm IGM}=&1.3\times 10^{-11} {\left 
(\alpha +3\over 4.5 \right )}^{-1/2}
\left({J_{-22}R_{100}\over a~\cos \theta}\right)^{1/2} T_{4.3}^{0.363} \\
\end{array}
\end{equation}
where $J\equiv J_{-22} 10^{-22}$ ergs s$^{-1}$ cm$^{-2}$ Hz$^{-1}$
sr$^{-1}$, $R=R_{100}100$ kpc, $T=T_{4.3}2\times 10^4$ K, $\alpha=1.5$
is the spectral slope of the ionizing UVB and B(z) is the number of
lines per unit $N_{HI}$ and unit redshift interval derived at given
redshifts from the observed spectra as in Giallongo et al. (1996). 
Thus, $dN/dz=B(z)\int N_{_{\rm HI}}^{-\beta} dN_{_{\rm HI}}$.
We include the contribution from Ly$\alpha$ lines of various column
densities on the basis of the known shape of the column density
distribution. For $N_{\rm HI}>10^{14}$ cm$^{-2}$ this distribution is
fairly steep, with $\beta _s\simeq 1.8$, and the contribution to
$\Omega_{\rm IGM}$ becomes progressively small.  For $N_{\rm
HI}<10^{14}$ cm$^{-2}$, $\beta _f\simeq 1.4$ and the contribution to
the mass density parameter increases slowly with decreasing $N_{\rm
HI}$.  We have adopted a lower $N_{\rm HI}$ cutoff of $10^{12}$
cm$^{-2}$ as derived from high resolution data (Webb et al. 1992,
Giallongo et al. 1995, Hu et al. 1995).

The cloud size and geometry are subject to substantial uncertainties.
An estimate of the characteristic size of absorbers is provided by the
statistical coincidence of absorption lines in closely separated
quasar pairs. Recent observations of the quasar pair 1343+264A/B at
$z\sim 1.8$ (Bechtold et al. 1994, Dinshaw et al. 1994) have shown
that the Ly$\alpha$ sizes are of the order of $R\sim 200$h$_{50}^{-1}$
kpc, much larger than previously thought.  Observations at lower
redshifts, $z\sim 0.5$, show even larger sizes, $R\magcir
360$h$_{50}^{-1}$ kpc, independently of the cloud structure or
geometry (Dinshaw et al. 1995). The mild redshift evolution implied by
these preliminary measurements is consistent with the expectations of
the standard CDM cosmological scenarios. As outlined by the numerical
simulations of Miralda-Escud\'e et al. (1996), although the gas moves
on average to regions of higher overdensity, the dominant effect for
the evolution of the average cloud properties appears to be the Hubble
expansion.

For these reasons, we have assumed a typical radius $R=200$h$_{50}^{-1}$
kpc at $z=1.8$ which evolves in time following the Hubble expansion as 
$R\propto (1+z)^{-1}$.

The ensuing cosmological baryon density due to Lyman-$\alpha$ clouds
is shown in Figure 2 from $z=0$ to $z\simeq 4$ as a function of cosmic
time.  The redshift dependence of the baryon density of photoionized
gas is associated with the product $(JR)^{1/2}dN/dz$ where the
dominant factor is the number density evolution of the clouds.

Two different evolutionary scenarios can be envisaged at this point,
depending on the sources of the ionizing background. In Figure 2{\it
a}~ $\Omega_{\rm IGM}(z)$ has been computed for a QSO-dominated
UVB. In Figure 2{\it b} both galaxies and QSOs contribute to the
metagalactic flux.  In the redshift interval $1.7<z<4.1$, the UVB
assumes the constant value $J_{912}\approx 5\times 10^{-22}$ ergs
cm$^{-2}$ s$^{-1}$ Hz$^{-1}$ sr$^{-1}$, as derived from the proximity
effect. The increase of the baryon density at higher redshifts is due
to the increase of the number density of Lyman-$\alpha$ lines, which
is partly compensated by the decrease of the average cloud sizes.
Lyman-$\alpha$ absorbers at $z\approx 4$ can easily account for all
the baryons in the universe predicted by nucleosynthesis, $\Omega_b
h_{50}^2=0.05\pm 1$ (Walker et al. 1991). Note that a large aspect
ratio, $a\magcir 10$, must be adopted to avoid $\Omega_{\rm IGM}$
values significantly in excess of the nucleosynthesis constraints, as
outlined by Rauch \& Haehnelt (1995).

The residual baryon density is already about 30\% at $z\sim 1.8$
(i.e., $t\sim 3$ Gyr). By contrast, for $z<1.7$ the number density of
Lyman-$\alpha$ clouds stays nearly constant and the evolution of the
density parameter depends mainly on the evolution of the ionizing
sources contributing to the UVB and of the cloud size. In any case, a
further decrease of $\Omega_{\rm IGM}$ is present in the redshift
interval $z=0.3-1.7$. This evolution is stronger in the case of a
QSO-dominated UVB where a value of $\Omega_{\rm IGM}\simeq 0.007$ is
found at $z=0.3$. In a galaxy-dominated UVB the evolution is slower
and $\Omega_{\rm IGM}\simeq 0.01$ at the same redshift. These values
are larger than the baryonic ``visible'' mass density of galaxies,
$\Omega_*=0.004e^{\pm 0.3}$ (e.g., Peebles 1993).

\begin{figure}
{~
\epsfxsize=250 pt
\epsffile{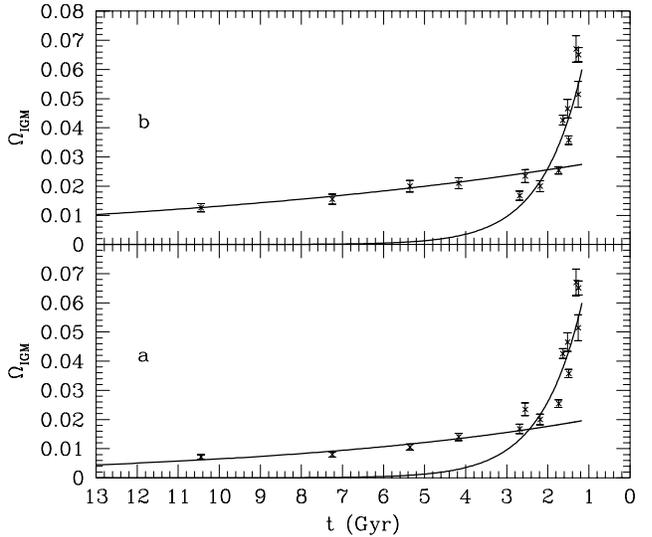}
}
\caption{ Mass density parameter of photoionized intergalactic gas as
a function of cosmic time. {\it a)} QSO-dominated UVB. {\it b)} Same
including a contribution from star-forming galaxies. Points are
derived from the Lyman-$\alpha$ sample described in the text. Two
exponential curves are shown simply for comparison (see sect. 4), with
$e$-folding timescales equal to 1 and 8 Gyr. Again, the galaxy
contribution has been computed assuming an escape fraction of ionizing
photons into the IGM of $\langle f_{esc} \rangle=15$\%.}
\label{fig2}
\end{figure}

\section{DISCUSSION}

We have presented here two main results. First, we have computed the
integrated emission of ionizing photons from the {\it observed}
population of star-forming galaxies at $z<1.3$, and shown that it may
equal or even exceed the overall QSO contribution to the UVB in this
redshift range if the mean fraction of UV photons which can escape
from individual galaxies into the IGM is not negligible.  We have
shown in \S~2 that $\langle f_{\rm esc} \rangle$ is constrained to be
$<20\%$, in order to satisfy the upper limit on the intensity of the
local UVB. If we adopt $\langle f_{\rm esc} \rangle \sim 15\%$ as a fiducial
value for the escape fraction, the ensuing UVB is found to evolve little in the
redshift interval between $z=0.4$ and $z=4$. 

We have also shown that the baryon fraction which is associated with
the photoionized Ly$\alpha$ clouds appears to decrease rapidly with
cosmic time.  At $z\sim 4$, $\Omega_{\rm IGM}$ may account for all of
the nucleosynthesis baryons of the universe. As shown in \S~3, the
mass density parameter depends only weakly on the intensity, $J_L$, of
the UVB, with $\Omega_{\rm IGM}\propto J^{1/2}$.  Hence tighter
constraints on the escape fraction from galaxies are unlikely to
change qualitatively the evolution of $\Omega_{\rm IGM}$ (see Fig. 2).
The dependence of $\Omega_{\rm IGM}$ on the typical cloud radius and
geometry is also relatively weak, $\Omega_{\rm IGM}\propto
(R/a)^{1/2}\equiv l^{1/2}$. Moreover, in CDM scenarios, lines with
$N_{\rm HI}<10^{15}$ cm$^{-2}$ have typical sizes which appear
essentially uncorrelated with the \HI column (Miralda-Escud\'e et
al. 1996). Thus, while the exact normalization of $\Omega_{\rm IGM}$
depends on these values, it is difficult to explain the strong
evolution inferred in the redshift interval $z=1.7-4$ as a simple
geometrical effect. Note that photoionization models and the
statistics of coincident absorptions in QSO pairs suggest a typical
cloud thickness in the range $2l\approx 100-300$ kpc, independently of
the assumed aspect ratio (Fang et al. 1996).  At $z\sim 4$, the baryon
content of the Lyman-$\alpha$ forest clouds may then be significant
both in the case of spherical or disk-like geometry.

We can parameterize the cosmological evolution of $\Omega_{\rm IGM}$
as an exponentially decreasing function with cosmic time, $\Omega_{\rm
IGM}\propto \exp (-t/\tau)$. As shown in Figure 2, this yields two
characteristic $e$-folding timescales, $\tau\approx 1$ Gyr for $t<3$
Gyr and $\tau\approx 8$ Gyr at later epochs.  Thus, roughly 60\% of
the photoionized gas in the universe ``disappears'' at early epochs
over a timescale which is much shorter that the Hubble time.  What is
the fate of these ``missing'' baryons? They are unlikely to fragment
into stars, as the baryonic visible mass density of galaxies today is
known to be only a small fraction of the value derived from
cosmological nucleosynthesis.  The decrease of $\Omega_{IGM}$ observed
in the Ly$\alpha$ line population must mainly reflects either a
dilution of the photoionized gas clouds in a general diffuse
intergalactic medium or an increase of the fraction of the IGM which
is collisionally ionized at temperatures $T\magcir 10^5$ K by
supernovae shocks.  If a strong evolution in the neutral column
density of clouds were present following the Hubble flow, it would
produce at $z=1.7$ lines ten times weaker ($\log N_{HI}\sim 11$) than
at $z\simeq 3.8$.  However, since the contribution to $\Omega_{IGM}$
increases slowly for decreasing $N_{HI}$ ($\Omega_{IGM} \propto
N_{HI}^{0.1}$ for $\beta \simeq 1.4$), lines ten times weaker in
$N_{HI}$ increase $\Omega$ only by $\sim 0.005$ when $z$ decreases
from $z\sim 3.8$ to $z\simeq 1.7$ (i.e. from $t\simeq 1$ Gyr to $t\sim
3$ Gyr). Thus the strong $\Omega$ evolution derived in the redshift
interval $z=1.7-4$ can not be mainly due to a decrease (with
decreasing redshift) of the lower cutoff of the $N_{HI}$ distribution
unless the sizes evolved much faster than assumed in this simple
model.

A plausible possibility appears to be the additional heating of the
absorbing gas at temperatures $T\sim 10^{5-6}$ K by the gravitational
accretion into progressively more massive halos, with higher velocity
dispersions, or by collisional ionization from supernovae winds.  The
easiest way for finding collisionally ionized, cosmologically
distributed material at $T\magcir 10^{5.5}$ K is to search for \OVI
absorption. \OVI is most prevalent at these temperatures, while \OVI
lines are stronger than those of \CIV for $T\magcir 10^6$ K. The
recent results of the first survey for \OVI 1032, 1038\AA\ absorption
lines in QSO spectra (Burles \& Tytler 1996) suggest the presence of a
substantial cosmological mass density of hot, collisionally ionized
gas at $\langle z\rangle =0.9$.  If the bulk heating were mainly due
to supernovae explosions in spheroidal systems, as suggested by recent
numerical simulations (e.g., Miralda-Escud\'e et al. 1996), the strong
evolution of $\Omega_{\rm IGM}$ observed between $z=2$ and $z=4$ could
be triggered by the star-formation activity in galaxies at high
redshift. Note that, while an $e$-folding timescale of only 1 Gyr is
much shorter than the decay time of star formation, $\approx 4-7$ Gyr,
characteristic of late type spirals, it is comparable with the decay
time of star formation characteristic of ``average'' quiescent
elliptical galaxies (e.g., Bruzual \& Charlot 1993). Various authors
have argued in favour of a significant contribution to the UVB by
star-forming galaxies (Bechtold et al.  1987; Miralda-Escud\'e \&
Ostriker 1990; Madau 1991; Madau \& Shull 1996).  Order-of-magnitude
arguments suggest that the overall kinetic energy released in
supernova explosions can be about one-third of the radiative UV power
(Miralda-Escud\'e \& Ostriker 1990). In this case a highly
photoionized IGM at $T\approx 10^{4.5}$ K may be quickly heated to
temperatures 10 times larger (Giroux \& Shapiro 1996). If this
scenario will be confirmed by new observational constraints on $R$ and
$J_L$ and by detailed theoretical models, then the evolution of the
Lyman-$\alpha$ forest clouds may be used as a probe of the cosmic star
formation rate as a function of time.

\bigskip
\noindent
{\bf ACKNOWLEDGMENTS}

\noindent
We thank the anonymous referee for criticism which improved the
clarity of this paper.

\bigskip
\noindent
{\bf REFERENCES}

\refer {Bahcall, J. N., et al. 1996, ApJ, 457, 19}

\refer {Bajtlik, S., Duncan, R. C., \& Ostriker, J. P. 1988, ApJ, 327, 570}

\refer {Barcons, X., Lanzetta, K. M., \& Webb, J. K. 1995, Nature, 376, 321}

\refer {Bechtold, J. 1994, ApJS, 91, 1}

\refer {Bechtold, J., Weymann, R. J., Lin, Z., \& Malkan, M. A. 1987, ApJ, 
	315, 180}

\refer {Bochkarev, N. G., \& Sunyaev, R. A. 1977, Soviet Astron., 21, 542}

\refer {Bruzual, G. \& Charlot, S. 1993, ApJ, 405, 538}

\refer {Cowie, L. L., Songaila, A., Kim, T.-S., \& Hu, E. M. 1995, AJ, 
	109, 1522}

\refer {Crampton, D., Le F\`evre, O., Lilly, S. J., \& Hammer, F. 1995, ApJ,
	455, 96}

\refer {Donahue, M., Aldering, G., \& Stocke, J. T. 1995, 450, L45}

\refer {Dove, J. B., \& Shull, J., M., 1994, ApJ, 430, 222}

\refer {Fang, Y., Duncan, R. C., Crotts, A. P. S., \& Bechtold, J. 1996,
	ApJ, 462, 77}

\refer {Ferrara, A, \& Field, G. B. 1994, ApJ, 423, 665}

\refer {Gallego, J., Zamorano, J., Arag{\'o}n-Salamanca, A., \& Rego, M.
       1955, ApJ, 455, L1}

\refer {Giallongo, E., Cristiani, S., D'Odorico, S., Fontana, A., \& Savaglio,
	S. 1996, ApJ, 466, 46}

\refer {Giroux, M., \& Shapiro, P. R. 1996, ApJS, 102, 191}

\refer {Haardt, F., \& Madau, P. 1996, ApJ, 461, 20}

\refer {Hurwitz, M., Jelinsky, P., Van Dyke Dixon, W. 1997, ApJ Letters,
	in press, astro-ph 9703041}

\refer {Leitherer, C., Ferguson, H. C., Heckman, T. M., Lowenthal, 
	J. D.  1995, ApJ, L19}

\refer {Lilly, S. J., Tresse, L., Hammer, F., Crampton, D., \& Le F\`evre, O.
	1995, ApJ, 455, 108}

\refer {Loveday, J., Peterson, B. A., Efstathiou, G., \& Maddox, S. J.  
	1992, ApJ, 390, 338}

\refer {Madau, P. 1991, ApJ, 376, L33}

\refer {Madau, P. 1992, ApJ, 389, L1}

\refer {Madau, P., Ferguson, H.~C., Dickinson, M.~E., Giavalisco, M., Steidel,
       C.~C., \& Fruchter, A. 1996, MNRAS, 283, 1388}

\refer {Madau, P., \& Shull, J. M.  1996, ApJ, 457, 551}

\refer {Miralda-Escud\'e, J., \& Ostriker, J. P.  1990, ApJ, 350, 1}

\refer {Miralda-Escud\'e, J., Cen, R., Ostriker, J. P., \& Rauch, M. 1996,
	ApJ, 471, 582}

\refer {Patel, K., \& Wilson, C. D. 1995a, ApJ, 451, 607}

\refer {Patel, K., \& Wilson, C. D. 1995b, ApJ, 453, 162}

\refer {Peebles, P. J. E. 1993, Principles of Physical Cosmology,
      (Princeton: Princeton University Press), p. 123}

\refer {Rauch, M., \& Haenelt, M. G. 1995, MNRAS, 275, 76}
 
\refer {Songaila, A., Cowie, L. L., \& Lilly, S. J. 1990, ApJ, 348, 371}

\refer {Steidel, C. C. 1995, in QSO Absorption Lines, Proc. ESO Workshop,
ed. G. Meylan (Heidelberg: Springer), p. 139}

\refer {Steidel, C. C., Giavalisco, M., Pettini, M., Dickinson, M., \&
	Adelberger, K. L. 1996, ApJ, in press}

\refer {Tytler, D., Fan, X.-M., Burles, S., Cottrell, L., Davis, C.,
Kirkman, D., \& Zuo, L.  1995, in QSO Absorption Lines, Proc. ESO Workshop,
ed. G. Meylan (Heidelberg: Springer), p. 289}

\refer {Vogel, S. N., Weymann, R., Rauch, M., \& Hamilton, T. 1995, ApJ, 
	441, 162}

\refer {Walker, T. P., Steigman, G., Schramm, D. N., Olive, K. A., \& Kang, H. 
	1991, ApJ, 376, 51}

\end{document}